\documentclass[referee]{aa} % for a referee version
\usepackage{graphicx}
\usepackage{txfonts}
\def\sbri{{\rm \;mag/\sqcup\hspace{-11pt}\;\sqcap ''}}

\begin{document}
   \title{Detection of Galaxies with Gaia }

   \author{R. E. de Souza 
          \inst{1}
          \and
          A. Krone-Martins
          \inst{3,2,1}
          \and
          S. dos Anjos
          \inst{1}
          \and
          C. Ducourant
          \inst{1,2}
          \and
          R. Teixeira
          \inst{1,2}
          }
   \titlerunning{Detection of galaxies with Gaia}

   \offprints{R. E. de Souza}

   \institute{Dept. Astronomia, IAG/USP, University of S\~ao Paulo,
   Rua do Mat\~ao 1226, 05508-900   S\~ao Paulo -SP, Brazil
   \and
   Observatoire Aquitain des Sciences de l'Univers, Laboratoire d’Astrophysique de Bordeaux, CNRS-UMR 5804, BP 89, 33271, Floirac Cedex, France
   \and
   SIM - Faculdade de Ci´encias da Universidade de Lisboa, Ed. C8, Campo Grande, 1749-016 Lisboa, Portugal}

   \date{Received: 27/01/2014 ; accepted: 16/04/2014}

% \abstract{}{}{}{}{} 
% 5 {} token are mandatory
 
  \abstract
     {}
   {Besides its major objective tuned to the detection of the stellar galactic population the Gaia mission experiment will also observe a large number of galaxies. In this work we intend to evaluate the number and the characteristics of the galaxies that will effectively pass the onboard selection algorithm of Gaia.}
   {The detection of objects in Gaia will be performed in a section of the focal plane known as the Sky Mapper. Taking into account the Video Processing Algorithm criterion of detection and considering the known light profiles of discs and bulges galaxies we assess the number and the type of extra-galactic objects that will be observed by Gaia.}
   {We show that the stellar disk population of galaxies will be very difficult to be observed. On the contrary the spheroidal component of elliptical galaxies and bulges having higher central surface brightness and steeper brightness profile will be more easy to be detected.   We estimate that most of the 20 000 elliptical population of nearby galaxies inside the local region up to 170 Mpc are in condition to be observed by Gaia. A similar number of bulges could also be observed although the low luminosity bulges should escape detection. About two thirds of the more distant objects up to 600 Mpc could also be detected increasing the total sample to half a million objects including ellipticals and bulges. The angular size of the detected objects will never exceed 4.72 arcsec which is the size of the largest transmitted windows.}   
   {An heterogeneous population of elliptical galaxies and bulges will be observable by Gaia. This nearby Universe sample of galaxies should constitute a very rich and interesting sample to study their structural properties and their distribution.}


   \maketitle
%
%________________________________________________________________

\section{Introduction}

The Gaia satellite experiment was launched in December 2013 as part of a mission to produce an all sky survey with unprecedented angular resolution in the spectral range 330-1000 nm. The selected objects will first pass through the Sky Mapper (SM) section of the focal plane responsible for filtering those objects that in the sequence should be observed in the Astrometric Field (AF). The expected software defined magnitude limit of the observable targets in the Gaia photometric system  is estimated to be $G=20$ mag for point source objects. The satellite is also equipped with two low resolution spectro-photometers operating in the blue spectral region 330-680 nm (BP) and in the red 640-1000 nm (RP) with a typical resolution of 3-27 nm per pixel for the BP photometer and 7-15 nm per pixel for RP. All the observed objects will have their spectral energy distribution evaluated by these instruments. Moreover an additional radial velocity spectrograph (RVS) instrument will measure the radial velocity in the Calcium triplet region (847-874 nm) for a fraction of the brightest objects with a resolution R=11500. The evaluation of the accurate position, distance and radial velocity for an enormous number of stellar objects will provide an extremely useful material to improve our understanding of our Galaxy (Perryman, \cite{Perryman2001}). Although the mission is focused on the detection of galactic stellar objects Gaia will also deliver a huge number of detections of non stellar objects such as quasars and particularly interesting data set of galaxies.  In the case of nearby galaxies in the local group it is expected  that their brightest stars will be also resolved providing an invaluable information for studying their stellar population and their dynamics. The more distant galaxies will be detected through their diffuse brightest regions mostly concentrated in their central regions. A natural question to ask in this scenario is related to the potential impact of the sample of detected non-resolved galaxies for our comprehension of this subject. 

It is clear that the sample of these non-resolved galaxies observations by Gaia will be limited to their very central and bright regions for nearby objects. In each passage of the telescope trough these galaxies, the on board satellite software will sample an small strip of pixels covering the information gathered from a region of $0.59$x$1.77$ arc sec (along scan vs. across scan) in the SM section. This region is used to estimate the magnitude of the object and to analyse if the object should be transferred to the Earth. An important point however is that the SM window send to Earth is larger and will cover an sky size of 4.72x2.12 arcsec. Since nowadays most of the morphological information occurring in the disks and bulges of nearby galaxies comes mainly from the observation of regions on much larger angular scales, its clear that the observed light distribution will be invaluable for morphological studies of the central region of galaxies. This sample will be a complementary set of observations to those that can be obtained from lower resolution Earth based facilities. However, it is important to mention that these data have reduced spatial resolution, reduced depth and increased noise since the integration time is $2.9$ s, instead of the usual $4.4$ s adopted for the AF field. Moreover, the SM window that reaches ground is composed of samples of 2x2 binned pixels for $G < 16$ mag. For $G > 16$ mag an on-board binning of four 2x2-pixel will be applied resulting in an increase of the total noise. As we will show in the next sections most of the unresolved galaxies should have $G > 16$ mag and therefore the SM Gaia observation of these objects will suffer from these limitations.

Due to its orbital characteristics the satellite will observe regularly the same region of the sky each time at different scanning directions. Therefore, it is possible to apply some numerical reconstruction method to obtain estimated two dimensional images for the observed objects, even though such approaches usually present severe reconstruction artefacts (see Dollet et al., \cite{Dollet2005} , Harrison, \cite{Harrison2011}, and references therein). Nevertheless the best angular resolution of Gaia's AF is 59 mas in the scanning direction and therefore we should expect that the its measurements of the central region of nearby galaxies will be available with a very good resolution, quite probably comparable with those available from the HST data and resulting in a larger dataset.

One possible way to evaluate the impact of Gaia observations is through the simulation of a complete distribution of objects in the Universe including solar system objects, galactic stellar sources and extragalactic sources like galaxies, QSOs and supernovae (Robin et al, \cite{Robin2011}). This could be a valuable approach for point-like objects, because the detection of such objects by the Gaia onboard software will be unbiased. Nonetheless, for extended sources, such as other galaxies, this large scale approach is infeasible, as it would required that images of all targets to be simulated and afterwards passed through the video processing algorithms. In Robin et al., 2011, simulation, a total of 38 million extragalactic objects were generated with integrated magnitudes $G \leq 20$ mag, obeying the observed luminosity functions. This figure could be considered as an upper limit of the possible number of detections by Gaia.

In the present work we explore a complementary view of the extra-galactic science that Gaia may produce focusing in the analysis of the possible structural components of galaxies that might be observed by Gaia. The two major components that we study are the disk and the bulge spheroidal components. Each of these two basic blocks have been extensively studied in the past decades having their own structural photometric properties well understood. Given the results discussed in the literature we then ask the question of what are the characteristic properties of these two components that might be more suited to be detected and selected for transmission to the ground selected for detection by the satellite.

In section 2 we present some general characteristics of the detection parameters adopted by Gaia that will be more relevant for the observation of galaxies. Section 3 discuss the implications for the detection of galaxies constituted by pure stellar disks assuming that those objects are dominated by an exponential surface brightness. Most of the stellar disks have a relatively low surface brightness distribution that will be hard to be detected by Gaia. In section 4 we analyse the detections of the spheroidal population located in ellipticals and bulges of spiral galaxies. This component is known to have a much higher central surface brightness and will probable be the major component to be detected by Gaia. In section 5 we describe the results of a numerical simulation of galaxies obeying the nearby observed brightness profiles. A population of these objects were distributed in a uniform space distribution and their simulated images pass trough a software emulating the same detection procedure adopted by Gaia. In section 6 we evaluate the effect of increasing the distance in the detection of more distant objects that should be affected by the fixed pixel aperture sampling and cosmological dimming. In section 7 we present our major conclusions.

\section{The Detection of Objects with Gaia}

The focal plane of Gaia used for astronomical observations is covered with 106 CCDs having 4500x1966 pixels of 10x30 $\mu$m   and the sky will be continuously monitored by the detectors. During its operation the reading of the CCD charges will be synchronised with the speed of the stars crossing the focal plane in a process known as time delay integration (TDI) with a TDI period of 982.8 $\mu$s. The resulting angular resolution will be 59 mas, along the scanning direction (AL), and 177 mas in the across scanning direction (AC). The observations of objects is planned to be performed in a two step process. The first pass will be conducted in the section called Sky Mapper (SM) having 14 CCDs where objects will be first detected and then selected for the higher resolution observations in the Astrometric Field (AF) section and also in the Blue and Red photometer sections (BP, RP). Depending on their detected magnitude in the SM section the selected objects with $G<17$ mag will be also observed in the radial velocity spectrograph (RVS) responsible for the radial velocity determination. 

The functioning of the SM section optimization procedure adopted in the Gaia mission is discussed by Provost et al ( \cite{Provost2007}) and  de Bruijne (\cite{Bruijne2014}). Here we will present only a very brief resume of this process with emphasis in our present subject of analysis. The first step of a detection will be applied by an algorithm running in the SM section where the video processing unit (VPU) build a synthetic pixel structure, called SM sample, electronically binning 2x2 of the physical Gaia pixels to provide a first lower resolution observation of each potential target in the sky. Each sample of the SM section will be continuously monitored by the video processing algorithm or VPA. Around each sample the VPA will define two concentric working windows with 3x3 and 5x5 samples. The detection algorithm will search the inner 3x3 windows looking for objects while the external region of the 5x5 windows will be used to determine the sky background. The signal of the external 16 samples of the 5x5 working window will be ranked and the fifth sample with the  lowest signal is adopted as the estimated background value. The total flux, $F$, is defined as the sum of the 3x3 working window corrected for the background value and applying a further correction to compensate for the predicted PSF losses of a point source object in the sampled window. The project evaluation is that in the average we should expect that typically a fraction of 80\%  the total stellar flux will be detected in the 3x3 sample working window and this figure is used to compensate for the PSF losses. The signal content of each sample ($S_{ij}$) will be added in the horizontal direction building three vectors $H_0$, $H_1$ and $H_2$, where $H_i=\sum S_{ij}$, as well as along the vertical direction $V_0$, $V_1$ and $V_2$. If $H_0\leq H_1>H_2$ and $V_0\leq V_1>V_2$ then there is a possible condition for the presence of an object detection in the central sample of this window. Then the two extreme horizontal and vertical vectors will be compared with the total flux to build the quantities $H_0/F$, $H_2/F$, $V_0/F$ and $V_2/F$ used as indicators of the PSF shape of each object. Based on the location of each detection in the planes $H_0/F$x$H_2/F$ and $V_0/F$x$V_2/F$ the VPA will classify the event either as a (1) Particle Prompt Event (PPE) mostly due to solar protons and galactic cosmic rays, a (2) Ripple region artificially created by the proximity of bright objects or as a (3) true celestial source generically called as a Star. By comparing a large body of simulations (de Bruijne , \cite{Bruijne2014}) the Gaia control mission developed a set of numerical relations defining exclusion regions to optimize the Gaia detection of stars.   

In case of positive detection a small region of the focal plane astrometric field (AF) around the object will be delivered to Earth receptors, together with the SM detection region, and will be available for further processing. This limitation in the size of the image send to Earth is necessary since the transmission of the whole mosaic of CCD frames is not feasible due to bandwidth communication restrictions of the satellite operation. The pieces of images of each detection of the same object observed in the AF section will be constituted by a set of  binned windows of the standard rectangular Gaia pixels measuring $177$ mas across the scanning direction (AC) by $59$ mas along the direction of the satellite motion (AL). An important issue due to the previously explained background determination procedure is that the decision of the on-board system to send each section of the image depends not only on their raw magnitude but also on the light gradient due to the illumination on nearby pixels. The expectation is that the adopted VPA detection algorithm will deliver all the stellar objects above the stipulated detection limit assuming a Gaussian PSF having $\sigma=(0.8-1.1){\rm x} 59$mas with a slight dependence on the stellar colour. For non-stellar images there will be obviously an additional bias against the detection of objects due to the SM adopted background correction (Krone-Martins et al, \cite{KroneMartins2013}).  

More distant galaxies will be detected by their brightest spots, in most cases restricted to their central regions. The exact properties of the final catalogue of detected objects will clearly depend on additional operational details of the satellite some of them yet under discussion. One interesting question consists in estimate which objects of the known nearby galactic population will have more chances to be detected by Gaia. This information is particularly useful in order to evaluate the probable scientific impact of the Gaia mission to our current understanding of galaxies. Two factors playing a major role in this context are the photometric system and the delivered satellite image resolution. The exact dimensions of the Astro-Field windows (AF) to be send to earth will depend on the detected flux in the SM section and for some objects the AF data it will be binned by the on-board video processing unit (VPU). Basically for brighter objects ($G<13$) the full pixel resolution will be preserved in the AF selected window. For fainter objects, which is the case for most galactic nuclei, it will be adopted  a binning procedure along the AC direction keeping the full resolution in the AL direction in order to preserve the final astrometric accuracy of the delivered images. In the AC direction the process of binning is designed to reduce the read-out noise by reading directly a large block of pixels. The final angular dimension of the transmitted window depends on the estimated object flux in the SM section as well as in the AF CCD sector column where the high resolution image of the object will be observed. For objects observed in the first AF CCD column (AF1) typical dimensions of the transmitted windows are $1.06$x$2.12$ arcsec for objects with $G<13$, $0.71$x$2.12$ arcsec for $13<G<16$ and $0.35$x$2.12$ arcsec for $G>16$. Therefore, since most galactic nuclei region should have $G>16$ mag, the expectation is that only the very central region of nearby galaxies will be available for ground processing of the AF observations. However, as remarked previously, for each object we will also have the SM sample region covering a size of 4.72x2.12 arcsec providing therefore a very useful material for ground processing of the detected images  even if these observations will suffer from an increase in the noise level as mentioned before. 

The photometric system adopted by Gaia is fully discussed in Jordi et al \cite{Jordi2010}. The main photometric information is obtained by the unfiltered light from 350 to 1000 nm defining the G magnitude system. Two additional passband will be also monitored by the blue (BP) and red photometers (RP) measuring the spectral energy distribution for each object trough a low dispersion optics. The spectral resolution depends on the wavelength. The BP photometer covers the spectral region 330-680 nm with a resolution 3-27 nm/pixel while the RP photometer will observe in the window 640-1000 nm with a resolution  3-27 nm/pixel. The third instrument will be used for stellar radial velocity determinations in the range 847-874 nm in the region of the Calcium II triplet. Only a fraction of the detected objects will be observed by this instrument. Using the RP data the VPU will estimate the magnitude, $G_{RVS}$, corresponding to the flux available in the RVS passband. Those objects with $G_{RVS}$ brighter than 17 mag will be observed and the expectation according to Katz et al., \cite{Katz2010} is that objects close to the limiting magnitude should present $S/N \simeq 1$ in a single transit. However is worth to mention that at the end-of-mission S/N integrated from all the transits will be higher than that, as each object is observed 40 times in average. 

The G photometric system is centred at $\lambda_0=6730$\AA \space with a passband approximately equal to $\Delta \lambda=4400$\AA. The two closest filters of the Johnson-Cousins system are V ($\lambda_0, \Delta \lambda =5510, 850 $\AA ) and R$_C$ ($\lambda_0, \Delta \lambda= 6470, 1570 $\AA). Several transformation equations are discussed by Jordi et al \cite{Jordi2010} and their accuracy depends on the color of the observed stars. In the case of normal galaxies the average color index of spirals and ellipticals are $B-V\simeq 0.4-1.0$ (Robertson \& Haynes \cite{RobertsHaynes1994}). The external region of spiral galaxies is dominated by an stellar disk with a mean color index  $B-V=0.5 \pm 0.1$ affected by relatively recent episodes of star formation. In the case of S0 and ellipticals, as well as in the classical bulges, the mean color is $B-V=0.8\pm 0.1$, due to an older stellar population with a mixture of objects having different ages and chemical compositions. Judging from these arguments the color transformation more adequate to be applied to galaxies are those developed for stars with $T_e>4500 K$ by  Jordi et al \cite{Jordi2010}

\begin{equation}
G-V_T=-0.0260-0,1767 (B_T-V_T) - 0.2980 (B_T-V_T)^2 + 0.1393 (B_T-V_T)^3
\label{Eq01}
\end{equation}

\noindent where the bands $B_T (\lambda_0,\Delta \lambda=4200, 710 )$\AA\space e $ V_T (\lambda_0,\Delta \lambda=5320, 980 )$\AA\space where defined from observations of stars in the Hipparcos catalogue. This system is close to the one of Johnson-Cousins $B(\lambda_0,\Delta \lambda=4410, 950 )$\AA, $ V (\lambda_0,\Delta \lambda=5510, 1570 )$\AA. According to Jordi et al (\cite{Jordi2010}) the mean expected residual by adopting this transformation is of the order of $0.03$ mag. The photometric transformation adopted in the Tycho catalogue, and valid in the interval $-0.2<(B-V)_T<1.8$, is $B-V=0.850 (B-V)_T$, with an error of 0.05 mag, and $V=V_T-0.090(B-V)_T$, with an error of 0.015 mag. Therefore an stellar object with a color equivalent to a typical spiral galaxy , $B-V=0.5$, should present a magnitude  $G=V-0.15$ in the Gaia photometric system.  In the case of a typical elliptical galaxy with a mean color index, $B-V=0.8$, the corresponding Gaia magnitude should be $G=V-0.26$.

As a crude first order estimation the adopted criteria for detection implies that the satellite should send to Earth objects with a limiting background corrected magnitude $G_{lim} = 20$ mag inside the 3x3 sample working window of the Sky Mapper Field (SM) section. We conclude that for close nearby objects the equivalent surface brightness detected by Gaia in the SM section, not corrected by background subtraction, should be approximately $\mu_G=G_{lim}+2.5\log(2{\rm x}3{\rm x}0.059 {\rm x}2{\rm x}3{\rm x}0.177)=18.9 \sbri$. As we can observe this surface brightness limit is relatively bright, when compared with surface brightness distribution of galaxies, indicating that the number of detected objects will be severely cut due to the small pixel size of Gaia's CCD and the adopted integration time defined by the TDI period. Moreover in this estimative we are assuming that the galaxy nuclear region will be detected with the same degree of efficiency as a point stellar object which is surely a very optimistic situation due to the SM adopted background criteria. A more realistic evaluation will  be done in the following sections showing  that a number of nearby galaxies will not be detect due to the light illumination effect on the neighbouring pixels of the nuclear region. On the other hand, as a compensating bonus, we should recognize that the SM detected objects having their data transmitted to Earth will have a very good sampling of the central regions with resolution comparable with those of the HST.

\section{Disks of Spiral galaxies}

The dominant structure in the outer regions of spiral galaxies is the rotationally supported stellar disk whose brightness profile along the major axis is normally described by the exponential profile approximation 

\begin{equation}
I_d(r) = I_{d0} \exp \left( -\frac{r}{r_d} \right)
\label{Eq02}
\end{equation}

\noindent where $I_{d0}$ represent the central surface brightness and $r_d$ is the scale length factor describing its gentle decreasing of light distribution (Freeman \cite{Freeman1970}). In terms of magnitudes this relation is equivalent to

\begin{equation}
\mu_d(r) = \mu_{d0} +1.08574 \frac{r}{r_d}
\label{Eq03}
\end{equation}

These expressions are considered as an useful description but are not meant to describe the full complexity of the true disk structure. Real galaxies show the presence of bars, rings, spiral arms and other distortions not included in such simplistic expressions. Nevertheless the exponential approximation capture the gross photometric structure of normal spiral galaxies dominated by the stellar disk population. One important effect is that the inclination of the disk along the line of sight causes a difference between the observed scale length profile along the major and minor axes and also affects the observed central surface brightness. In the approximation of a thin circular disk the observed isophotal contours have an elliptical shape and this projection effect can be corrected by evaluating the face-on central brightness as

\begin{equation}
\mu_{d0}(face-on) = \mu_{d0}(obs) -2.5 Log(b/a)
\label{Eq04}
\end{equation}

\noindent where $b$ and $a$ are the dimensions of the minor and major axis respectively.

In its paper Freeman showed using photographic observations in the blue band that exponential disks of galaxies tend to present an extrapolated central surface brightness concentrated at $\mu_{0B} \simeq 21.65 \sbri$ an statement usually known as the Freeman's law. Posterior analysis have extended this study of the photometric disk structure to a large number of objects showing that the central surface brightness is almost independent of the Hubble class but is not really constant presenting a relatively large dispersion covering the interval  $\mu_{0B} =(17-24) \sbri$. Moreover there is a loose correlation between $\mu_{0d}$ and $r_d$ indicating however the fact that the disk luminosities are distributed over a relatively small interval (Kent \cite{Kent1985}).

The apparent luminosity of an exponential disk inside a radial distance $r$ is obtained by integrating equation \ref{Eq02} over the elliptical image of a thin disk seen in projection. The resulting expression for the total luminosity of a face-on disk is 

\begin{equation}
L_T=2\pi I_{0d} q r_d^2 
\label{Eq05}
\end{equation}

\noindent where $q=b/a$ is the axial ratio. This expression implies that the absolute magnitude for a face-on object can be expressed as

 \begin{equation} 
M_{abs}=-38.5676+\mu_{0d} -5\log r_d(kpc)
\label{Eq06}
\end{equation}

Another useful quantity is the effective radius ($r_{e}$) indicating the region where the integrated luminosity of the disk drops to half of its total value, 

\begin{equation}
r_{e}=1.87635 r_d
\label{Eq07}
\end{equation}

\noindent resulting that the effective surface brightness is given by $\mu_e=\mu_{0d}+2.03723$.

\begin{figure*}
 \centering
 \includegraphics[width=\textwidth]{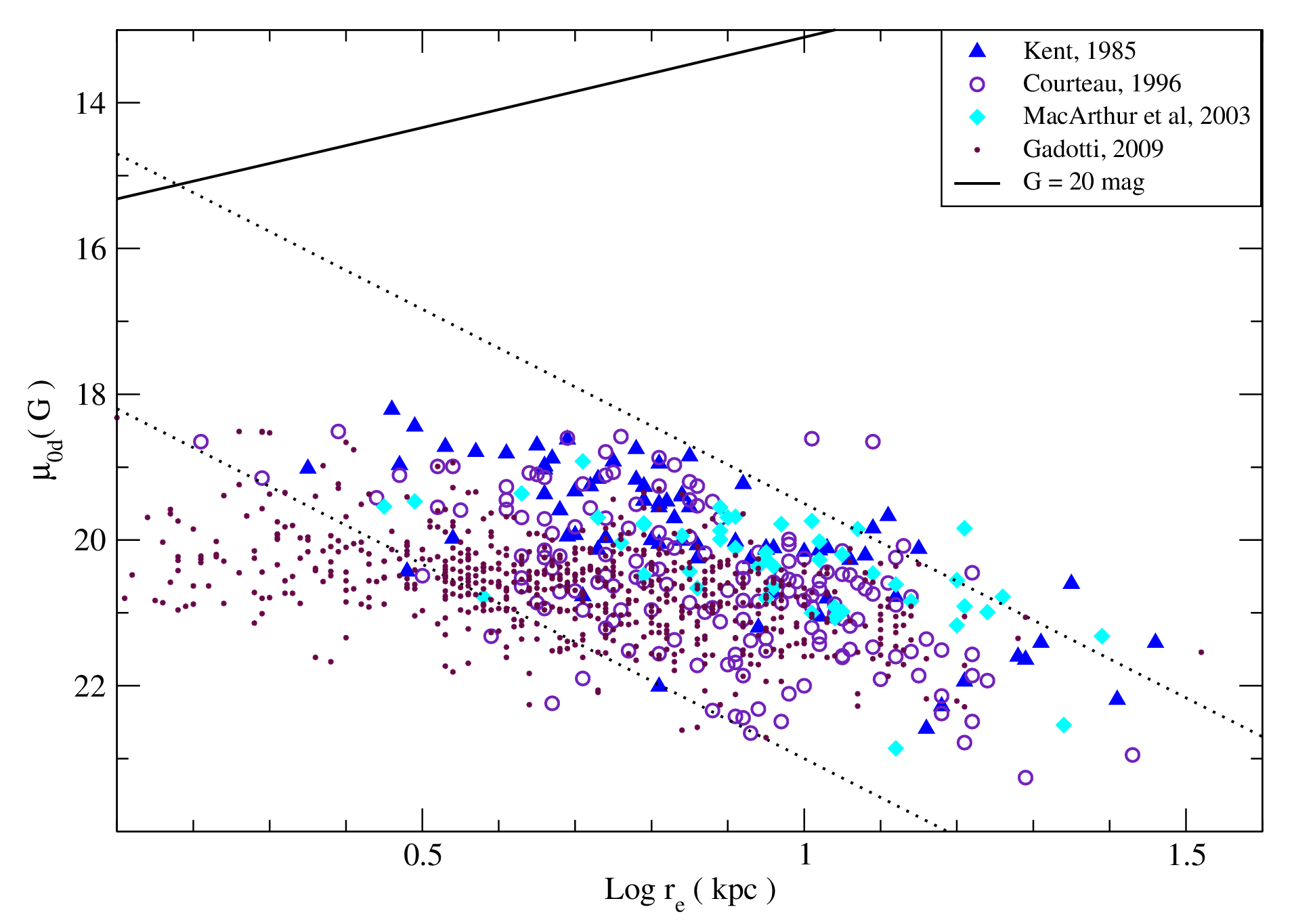}
 \caption{Central brightness distribution of spiral disk in the samples of Kent \cite{Kent1985}, Courteau, \cite{Courteau1996}, MacArthur et al, \cite{MacArthur2003} and Gadotti, \cite{Gadotti2009}. The two dashed lines corresponds to the limiting region containing spiral disks having absolute G magnitude of -18 and -22 mag.  The solid inclined line corresponds to the limiting surface brightness magnitude for object detection in the SM section, at a fiducial distance of 10 Mpc. Objects bellow that line have a central surface brightness too low to be detected by Gaia. As we can observe none of the observed pure exponential disks are in condition to be sampled by Gaia}\label{Mu0ReDisk}
\end{figure*}

In figure \ref{Mu0ReDisk} we present an overview of the structural properties of the observed exponential disks  derived from three selected samples extracted from the literature. The sample used by Kent (\cite{Kent1985}) have 105 objects observed in the r band of the uvgr Thuan \& Gunn (\cite{ThuanGunn1976}) system. The large majority of objects in this sample are spirals and a few elliptical galaxies (11 objects) were also included. In order to bring these observations to the G photometric system of Gaia we have first transformed the data to the UBV Johnson-Cousin system using the equations  quoted in Kent (\cite{Kent1985b}). Then we have used a mean estimated colour $B-V=0.5$ to convert the data to the G system using the transformations discussed in the previous section. The second sample was extracted from Courteau (\cite{Courteau1996}) derived from observations in the r Spinrad Lick filter and we have adopted the calibration quoted in the paper to match the system used by Kent (\cite{Kent1985}) and then applying the same transformation used before to the G system. We also have applied to the Courteau (\cite{Courteau1996}) data the same face-on correction adopted by Kent (\cite{Kent1985}). The sample of MacArthur et al (\cite{MacArthur2003}) was observed in the BVRH system easing the transformation to the G system from the results directly observed in the V band. The data in Gadotti ( \cite{Gadotti2009} ) is composed by nearly 1000 galaxies extracted from the Sloan Digital Survey and analysed with the program BUDDA (de Souza et al, \cite{BUDDA}) which is able to fit a Bulge+Disk+Bar model directly to the observed images. 

A first word of caution should be made due to the uncertainty in comparing the photometric systems used in these studies. Moreover the process of deriving the disk structural parameters are also different. The analysis of Kent (\cite{Kent1985}) was done through a fit of the minor and major axis brightness profiles to a combination of de Vaucouleurs $r^{1/4}$ bulge plus an exponential disk profile. Courteau (\cite{Courteau1996}) have adopted an elliptical approximation describing the isophotal levels in order to obtain the major axis light profiles which were fitted to a composite model with an exponential disk plus an $r^{1/4}$ bulge model. MacArthur et al (\cite{MacArthur2003}) have used a combination of an exponential disk with a S\'ersic $r^{1/n}$ law to describe the bulges. Nevertheless we can see from figure \ref{Mu0ReDisk}  that the spiral disks of these three samples consistently occupy approximately the same region of the $\mu_{0d} - Log {r_{e}}$ diagram with an small offset probably due to a combination of the different approaches adopted to determine the structural parameters. The two traced  lines indicates the prediction for an exponential disk having absolute G magnitude of $-18$ and $-22$ and we can see that they are approximately consistent with the observed trend of the empirical points as pointed out by Kent (\cite{Kent1985}).  

The continuous line in figure \ref{Mu0ReDisk} correspond to the SM detection limit G=20 mag computed by assuming a fiducial distance of 10 Mpc representative of a close nearby spiral disk. For a given combination of the parameters $\mu_{0d}$ and ${r_{e}({\rm kpc})}$ we correct for the distance effect in the pixel illumination and estimate the surface brightness distribution using equation \ref{Eq02}. Then we use a numerical interpolation routine dividing each Gaia physical pixel into a 30x30 mesh to compute the total flux contribution per pixel. The estimated fluxes were evaluated as they will be observed in the 5x5 SM sample working windows applying the VPA Gaia detection criteria presented in section 1.  We observe from the continuous line in figure \ref{Mu0ReDisk} that the surface brightness detection limit is relatively close to our crude estimation presented on the end of the previous section for objects with small effective radius but it also shows a consistent trend to be brighter for objects with large effective radius. This effect arise because a lower effective radius implies in a steeper profile with a larger fraction of the disk flux concentrated in the central pixel and as a consequence the evaluated SM background becomes lower when compared with the central surface brightness of the galaxy. On the contrary in disks with larger effective radius the light is more diffusely distributed and only those objects with extremely high central surface brightness could be detected. We can further verify that the observed spiral disks in the local Universe showing the highest central surface brightness are located in the region $\mu_{0d}(G)\simeq 18 \sbri$ well below the detection limit of the Gaia system. We should also point that by increasing the distance the estimated apparent magnitude in a given SM working window of Gaia is also increased due to a combination effect of the background subtraction procedure and to the fixed size of each pixel requiring therefore higher levels of surface brightness to reach the stipulated detection limit. Therefore we are forced to conclude that the observations of the Gaia satellite will not be able to examine the properties of the spiral disks observed in the local universe. Even if we consider that the detected images by Gaia will be downloaded including the surrounded pixels within 3 magnitude level only the very bright nearby disks will have any chance to be sampled. The only possible exception will be those objects having a bright non-stellar nuclei as an AGN and objects with strong nuclear star forming regions. 

\section{Spheroidal Components}

One of the most commonly adopted approximations to describe the luminosity distribution of spheroidal components, including bulges of spirals and elliptical galaxies, is the relation proposed by de Vaucouleurs (\cite{deVauc1948})

\begin{equation}
I(r)=I_{e} \exp\left\lbrace-7.66925\left[\left( \frac{r}{r_{e}}\right)^{1/4}-1  \right] \right\rbrace
\label{Eq08}
\end{equation}

\noindent where the constant $k=7.66925$ is defined in such a way that $r_{e}$ represent the effective radius containing  half of the total luminosity and $I_{e}$ is the corresponding surface brightness at the effective radius. For obvious reasons this relation is also known as the  $r^{1/4}$ profile. As a consequence the central surface brightness of the spheroidal component is

\begin{equation}
I_0=I_{e} e^{7.66925}=2141.47 I_{e}
\label{Eq09}
\end{equation}

\noindent or $\mu_{0}=\mu_{e}-8.327$ indicating that for a given effective surface brightness the spheroidal structure tend to present a much larger central surface brightness than the one observed in the exponential disks easing therefore its detection by Gaia. The total luminosity of the  $r^{1/4}$ profile derived from this expression is

\begin{equation}
L_T= \frac{8!\exp (7.66925)}{7.66925^8} \pi r_{e}^2q I_{e}=7.21457\pi r_{e}^2qI_{e}
\label{Eq10}
\end{equation}

\noindent where $q=b/a$ represent the axial ratio of the adopted structure of concentric elliptical isophotal levels. The absolute magnitude derived for this profile is

\begin{equation}
M_{abs}=-39.961 +\mu_{e}-5\log r_{e}(kpc) -2.5 \log q
\label{Eq11}
\end{equation}

\noindent where the effective radius is expressed in Kpc.

For some objects the $r^{1/4}$ profile gives a good empirical representation of the data. But in several cases this profile is clearly not adequate to represent the whole dynamical range of the observed surface brightness distribution.  In the recent years a preference have been established to describe the spheroidal component by using the S\'ersic (\cite{Sersic1969}) $1/n$ law  which is a generalization of the de Vaucouleur's law. 
 
\begin{equation}
I(r)=I_{e} \exp\left\lbrace  -b_n\left[\left( \frac{r}{r_{e}}\right)^{1/n}-1  \right] \right\rbrace
\label{Eq12}
\end{equation}

By adopting $n=4$ the normalization constant becomes $b_n=7.66925$ and the S\'ersic law reproduce the de Vaucouleur's law. For $n<4$ the light distribution has a softly less concentrated profile. In particular when $n=1$ the constant is $b_n=1.67834699$  and the profile is identical to the one used to describe the exponential disk. For $n>4$ the S\'ersic profile has a hard highly concentrated brightness distribution. The constant $b_n$ is determined by the condition that $r_e$ and $I_e$ are the effective radius and surface brightness respectively and one useful approximation to evaluate this constant was derived by Ciotti \& Bertin (\cite{Ciotti1999}).
 
\begin{equation}
b_n \simeq 2n-\frac{1}{3}+\frac{k_1}{n}+\frac{k_2}{n^2}+ \frac{k_3}{n^3}-\frac{k_4}{n^4} + ...
\label{Eq13}
\end{equation}

\noindent where $k_1=4/405$, $k_2=46/25515$, $k_3=131/1148175$ and $k_4=2194697/30690717750$.

The central surface brightness of the S\'ersic profile is 

\begin{equation}
I_0=I_{e} e^{b_n}
\label{Eq14}
\end{equation}

\noindent indicating that $\mu_0=\mu_{e}-1.0857b_n$ and we can observe that the central surface brightness depends on the value adopted for the exponent $n$. In the regime $n>4$ the central brightness becomes increasingly larger implying in a brighter central region for a given effective brightness. The analogous relation for the asymptotic total luminosity is 

\begin{equation}
L_T=  \frac{2\pi n}{b_n^{2n}} \Gamma(2n) I_0 r_{e}^2q
\label{Eq15}
\end{equation} 

\begin{figure*}
 \centering
 \includegraphics[width=\textwidth]{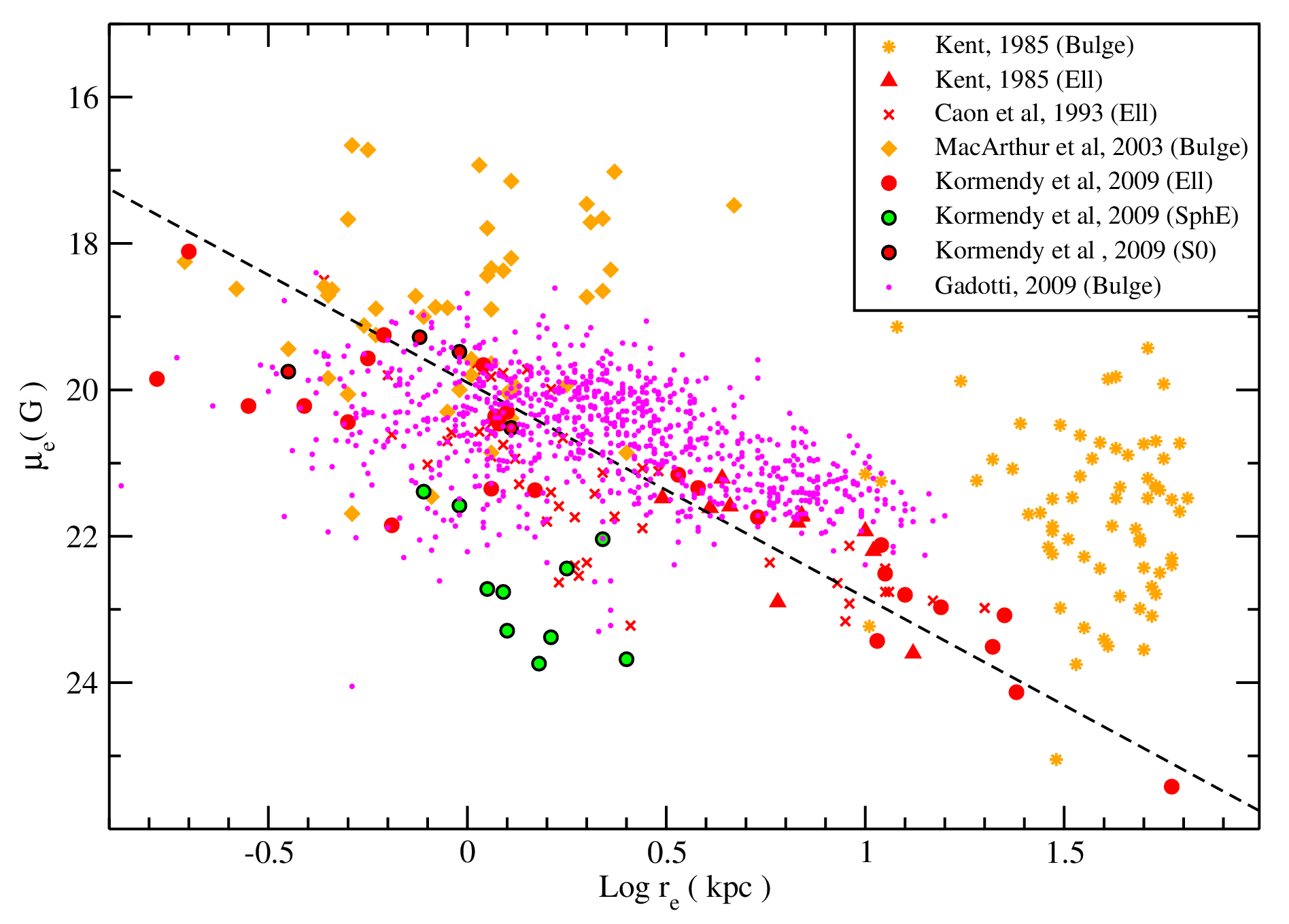}
 \caption{Correlation between the effective brightness and effective radius observed in nearby spheroids. The orange circles and red triangles represent respectively the bulges and elliptical from Kent \cite{Kent1985}. The red crosses are the Virgo cluster ellipticals from Caon et al \cite{Caon1993}. The orange diamonds are the bulge data from  MacArthur et al, \cite{MacArthur2003}. The red circles are the ellipticals from Kormendy et al \cite{Kormendy2009}. The small pink circles represent the sample of bulges studied by Gadotti, \cite{Gadotti2009}. The dashed line represent the Hamabe \& Kormendy, \cite{HamabeKormendy1987} empirical relation.   }\label{MueReSph}
\end{figure*}

From the image decomposition analysis of several objects it emerges the existence of a correlation between the effective brightness and effective radius of elliptical and bulges generally known as the Kormendy relation (Hamabe \& Kormendy, \cite{HamabeKormendy1987}). This correlation, represented by the dashed line in figure \ref{MueReSph}, indicates that there is a fundamental link, established during the formation process, between the structural parameters describing the light distribution of spheroids.

\begin{equation}
\mu_{e}(V)= 2.94 {\rm Log }\; r_{e}(kpc) +19.48
\label{Eq16}
\end{equation}

Using the de Vaucouleur's profile we can observe that the Kormendy relation together with equation \ref{Eq11}, or equation \ref{Eq15} for the corresponding S\'ersic profile,  implies that the more luminous objects are located in the extreme lower right region of figure \ref{MueReSph} having therefore a larger effective radius and also a lower effective surface brightness. The red triangles and orange circles in that figure comes from the analysis of Kent (\cite{Kent1985}) who adopted the de Vaucouleur's law to describe the spheroidal component of bulges and ellipticals. The sample of spirals investigated by Kent are spread over the whole morphological types and several objects have S0-Sa early type morphology.  The red crosses represent the data from Caon et al (\cite{Caon1993}) observations of 52 elliptical and S0 galaxies in the Virgo cluster fitted with the S\'ersic law. These authors have demonstrated that the $r^{1/n}$ profile, with $1<n<15$ give a more unbiased and accurate representation of the observed brightness profile. The vast majority of these objects have $n=3.8\pm 5$ with no clear distinction between ellipticals and S0. The orange diamonds indicate the position of the bulges in the sample of MacArthur et al (\cite{MacArthur2003}) 121 spiral galaxies most of them having morphological types later than Sb. According to these authors the adopted exponent of the bulge $r^{1/n}$ profile is observed to be restricted to the region $n<3$ peaked around $n=1 \pm 0.4$ close to the pure exponential solution. Therefore the light distribution of bulges in this late type sample is much less concentrated than the results found for S0 galaxies in Kent (\cite{Kent1985}) and Caon et al (\cite{Caon1993}). Moreover spirals in the sample of MacArthur et al, (\cite{MacArthur2003}) tend to present a correlation between the bulge effective radius and the disk scale length $<r_{eb}/r_d>=0.22 \pm 0.09$ indicating that the bulges of late type spirals are more affected by the presence of the stellar disk. The sample of bulges of spirals extracted from the SDSS and studied by Gadotti ( \cite{Gadotti2009}), indicated by the pink dots, obeys approximately the same relation observed in elliptical galaxies with a difference in the angular coefficient. Finally the circles are the spheroidal objects with high HST resolution nuclear images discussed in the sample of Kormendy et al \cite{Kormendy2009}. Their analysis indicate that $2<n<15$ concentrated at $n=2-4$ showing a different behaviour between ellipticals and S0 (red circles), low luminosity spheroidal ellipticals (green circles). We can observe that despite of the difference in photometric resolution and method of analysis there is a relatively good agreement between these samples in the description of the overall photometric properties of ellipticals and bulges. 

The observed dispersion around the mean trend of the Kormendy relation is real and results from the fact that this relation can be interpreted as a projection of a deeper relationship describing the fundamental plane of spheroids (Djorgovski et al, \cite{Djorgovski1996}). One additional interesting feature that we can observe is that for a given effective radius the surface  effective brightness of bulges tend to be brighter than the corresponding elliptical. This trend is more evident both in the samples of Kent (\cite{Kent1985}) and MacArthur et al (\cite{MacArthur2003}) with the difference that the bulges of early type spirals seem to be smaller and brighter than the corresponding bulges of late type galaxies. However, in the sample of Gadotti ( \cite{Gadotti2009}), where the photometric components are simultaneously derived from a direct fit of the image, this effect is more subtle. Therefore it might be possible that the differences between the Kormendy relation in bulges and ellipticals is partially affected by the adopted process used to decompose the relative contributions of disks and bulges in spiral galaxies.

\begin{figure*}
 \centering
 \includegraphics[width=\textwidth]{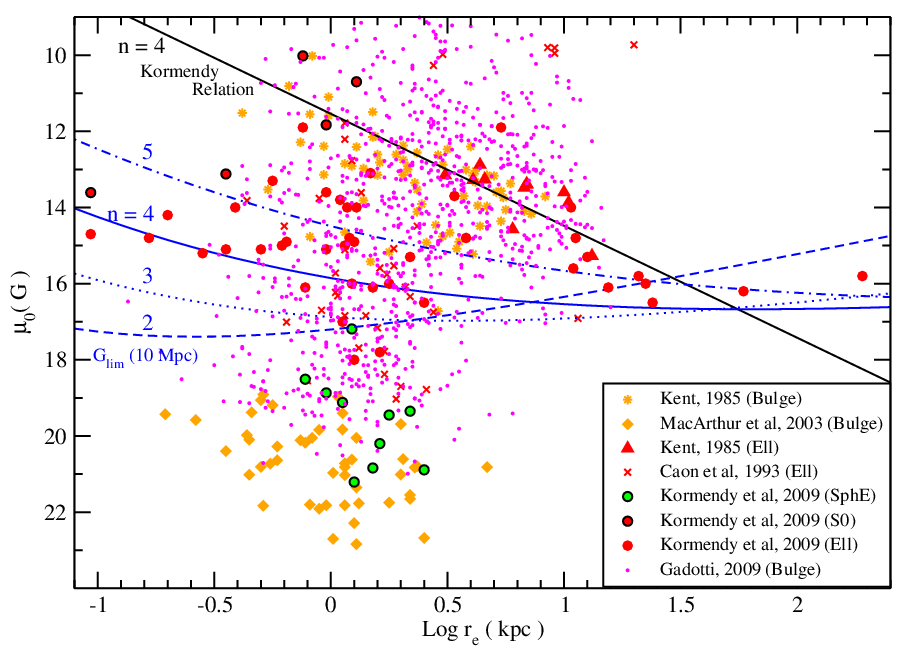}
 \caption{Central brightness distribution versus effective radius of spheroids of the same data sources used in figure \ref{MueReSph}. The black line indicate the predicted position of brightness profiles obeying the Kormendy relation for an index $n=4$ of the S\'ersic profile. The vertical spread of the data is due to variations of the index $n$ observed in the observed profiles. Objects with steeper profiles ($n>4$) are located above that line while objects with softer profiles ($n<4$) populate the lower region. The blue lines correspond to the limiting surface brightness of S\'ersic spheroids at a fiducial distance equal to 10 Mpc having $G=20$ mag in the Gaia SM section. Objects located above the limiting line are in conditions to be detected and depending on the S\'ersic index ($n$) the location of the limit of detection will be different. For $n>4$ the predicted location of most the objects, according to the Kormendy relation, is situated above the limiting curve and therefore they should be detected by Gaia.  On the contrary, spheroids having softer profiles, as the sample of the MacArthur et al (\cite{MacArthur2003}) bulges tend to be located below the Gaia detection limit and should not be detected. }\label{Mu0ReSph}
\end{figure*} 

In figure \ref{Mu0ReSph} we present the spheroidal data in the $\mu_0$x$log r_e$  plane which is more relevant in the discussion of Gaia observations. The black continuous line indicate the predicted position of objects obeying the Kormendy relation for $n=4$ brightness profiles approximately following the trend observed in the samples of Kent (\cite{Kent1985}) and Caon et al (\cite{Caon1993}). We can still observe as mentioned in the discussion of figure \ref{MueReSph} that for a fixed value of $n$ the larger and more luminous objects tend to present a lower central surface brightness along that line. The region above that line is populated by objects having a more steeper brightness profile with $n>4$ while the inferior region is richer in objects having a more softly ($n<4$) profiles. A variation $\Delta n=1$ in the index of the S\'ersic profile will displace the position of the Kormendy relation by $\Delta \mu_0=2.17$ mag increasing the vertical spread seen in this diagram. Most of the bulges observed in the sample of late spirals of MacArthur et al (\cite{MacArthur2003}) are located in the lower ($n \simeq 1$) region displaced by approximately 6.5 mag relative to the $n=4$ kormendy relation.

The blue lines corresponding to $n=$2, 3, 4 and 5 in figure \ref{Mu0ReSph} represent the limiting surface brightness corresponding to the VPA detection criteria adopted by the Gaia Sky Mapper to identify a G=20 mag object located at a fiducial distance $d=$ 10 Mpc using the same numerical integration routine as explained in section 2.  At that distance we can observe that for $n>2$ most of the normal spheroids, including a large fraction of bulges, are located in the region of detection of the Gaia satellite with the possible exception of a few large and bright objects. For $n<2$ the observed central brightness is lower and considering the additional difficulty due to the background subtraction it results that a larger number of bulges and ellipticals having soft brightness profile tend to escape the limit of detection. In particular the bulges of late spirals observed by MacArthur et al ( \cite{MacArthur2003}) some of the spheroidal galaxies in the sample of  Kormendy et al ( \cite{Kormendy2009}) and ellipticals with $n<2$ quite probably will not be detected by Gaia. Nevertheless we can observe from the large homogeneous sample of Gadotti ( \cite{Gadotti2009}), representative of the whole distribution of morphological types, that Gaia should be able to observe a large fraction of local bulges having relatively steeper brightness profiles.

Another point of concern is that the extrapolation of the S\'ersic profile to the central region below 1 arcsec  indicate some extremely high central surface brightness that are not realistic. This effect was already recognized by Caon et al (\cite{Caon1993}) and confirmed from data of HST high resolution images of  Kormendy et al ( \cite{Kormendy2009}) among others. In their very central regions ellipticals show a dichotomy with some objects presenting an inner core where the surface brightness is lower than predicted by the S\'ersic profile.  In other objects the observed profile show a cuspy behaviour where the surface brightness is significantly higher than the S\'ersic profile. According to Kormendy et al (\cite{Kormendy2009}) there is a trend in the sense that bright ellipticals tend to show a core profile while lower luminosity objects with $-21.54<M_V<-15.53$ tend to present an extra light cuspy  central region. Moreover no object exist in this whole sample with a  central surface brightness in excess of $10 \sbri$. Therefore the very high brightness inferred from $n>4$ profiles should be treated with high caution. In fact we could expect that the observations of Gaia should be of great help in clarifying this effect by the observation of a large number of objects.

\begin{figure*}
 \centering
 \includegraphics[width=\textwidth]{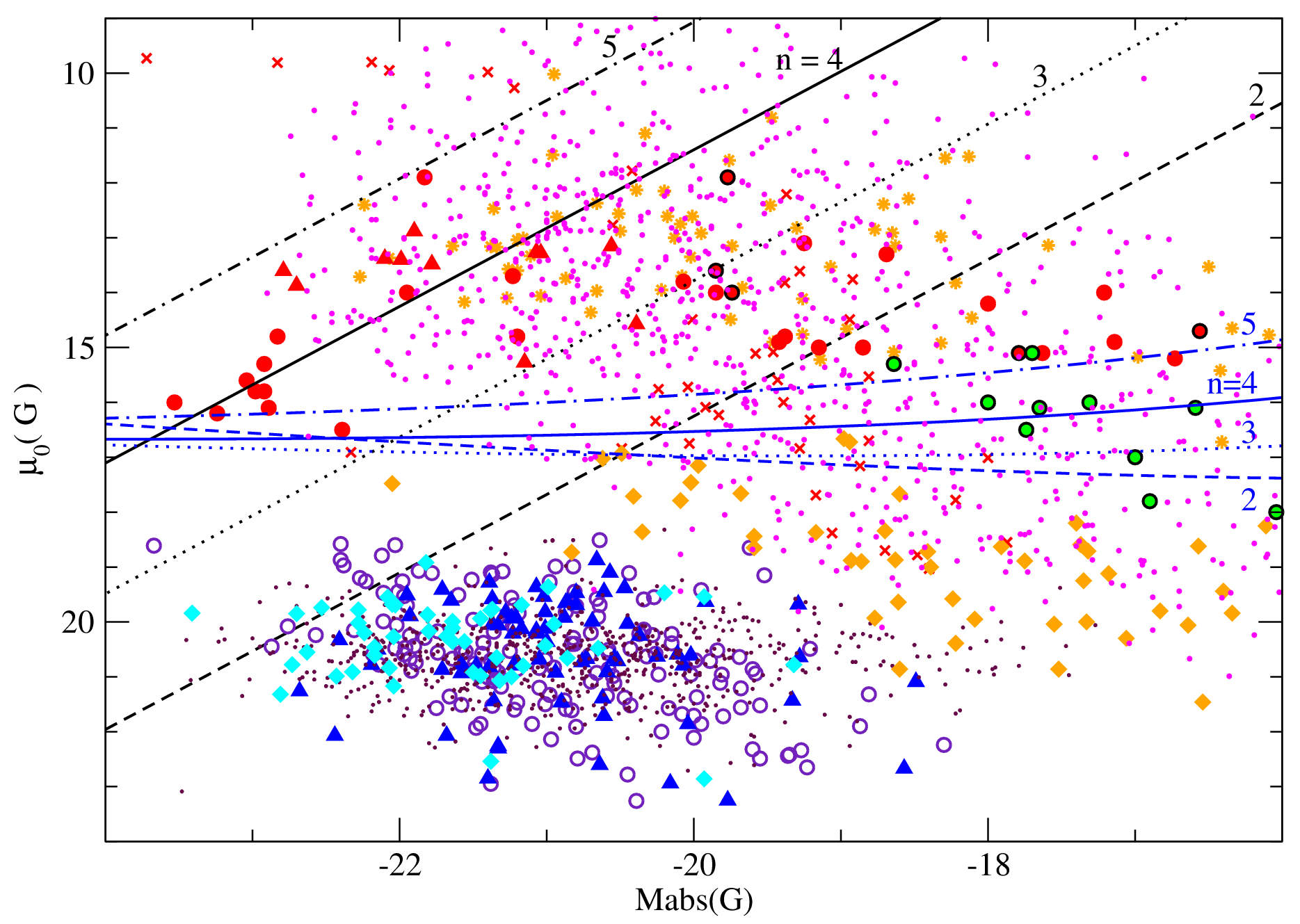}
 \caption{Relation between the central brightness and absolute magnitude distribution for spheroids and stellar disks. The symbols are the same as in figures \ref{Mu0ReDisk} and \ref{MueReSph}. Local stellar disks and bulges of late type spirals are located below the Gaia detection limit for a fiducial distance of 10 Mpc and should not be detected. However the vast majority of normal ellipticals and classical bulges have a sufficient high central surface brightness to detected.}  \label{Mu0Mabs}
\end{figure*}

The detection limit imposed by Gaia can be also examined using all objects, including disks and spheroids, in a plot of the central surface brightness against the total absolute magnitude of each component as presented in figure \ref{Mu0Mabs}. In this diagram the disks of spiral galaxies are clearly segregated to a relatively small region well below the detection limit of Gaia as we have already concluded. Directly above this region we can see the broad inclined strip domain occupied by elliptical galaxies having S\'ersic exponent in the interval $n=3-5$. The black lines indicate the predicted locus of spheroidal components obeying the Kormendy relation. The blue lines correspond to the same Gaia limiting region for detection in the SM section at a fiducial distance of 10Mpc seen in figure  \ref{Mu0ReSph}. Objects above that line are in conditions to be detected by Gaia. We can observe that objects similar to the bright luminous  elliptical galaxies with $M_G>-22$ studied by Kormendy et al (\cite{Kormendy2009}) and having high definition images from HST should be located close to the Gaia's detection limit. As mentioned before those very bright ellipticals tend to present a larger effective radius and as a consequence of the Kormendy relation their effective surface brightness is lower, as well as their central surface brightness. Except for these few very luminous objects the vast majority of ellipticals are located well inside the detection region of Gaia. The situation of bulges is less clear. The bulges of early type spirals observed by Kent (\cite{Kent1985}) and the few S0 objects studied by Kormendy et al (\cite{Kormendy2009}) are consistently close to the region occupied by ellipticals also inside the detection limit of Gaia. However a large fraction of the bulges of late type spirals in the sample of MacArthur et al, \cite{MacArthur2003} have a more soft brightness profile with $n<2$ and are well below the limit of detection. The same situation is observed in the case of some spheroidal ellipticals that should escape to Gaia detection. In the sample of  Gadotti, \cite{Gadotti2009} we can see that only those low luminosity bulges with low surface central surface brightness will escape detection but the vast majority of normal bulges are well inside the detection region. 

\section{Numerical simulations} 

\begin{figure*}
 \centering
 \includegraphics[width=0.99\textwidth]{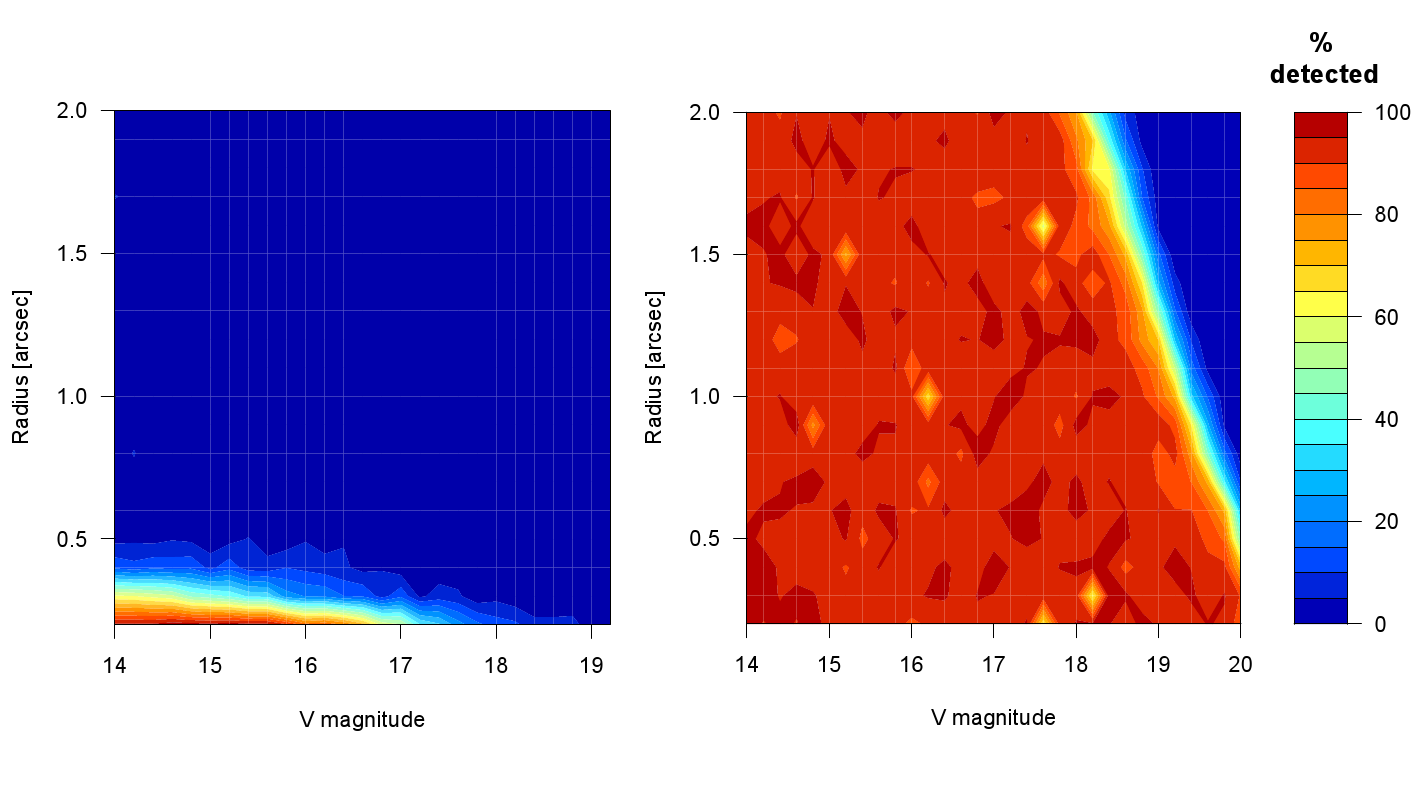}
 \caption{Efficiency detection maps of simulated brightness profiles at the Sky Mappers produced by the Gaia VPA Prototype implemented in GIBIS. In the left, exponential profiles. In the right, de Vaucouleurs profiles. }\label{GalaxySMdetectionsVPA}
\end{figure*} 

An exact scenario of the detections of extended objects by Gaia will only be available after the first months of mission, during a phase called Initial In Orbit Calibration. At that phase, the satellite will continuously observe a region around the ecliptic pole, and thus the physical characteristics of the objects detected and transmitted to the Earth may be determined. However, to assess if the scenario described in this paper is supported by the prototype of the onboard video processing algorithm, we can already perform a comparison with results from realistic simulations of the instrument. 

To do so, we performed fully numerical simulations of galaxy images in the Gaia focal plane and of their detection. These simulations adopted the Gaia Instrument and Basic Image Simulator, or GIBIS (Babusiaux et al.  \cite{Babusiaux2005}, Babusiaux et al.  \cite{Babusiaux2011} ), and the onboard video processing algorithm prototype implemented in this simulator. 

The simulations were performed at the $(l,b)_{gal} = (40.0^\circ, 52.0^\circ)$ direction. These coordinates were chosen for statistical reasons, as the GIBIS scanning law predicts a large amount of observations around it. As the simulated profiles are symmetric, any other positions on the sky presents similar results, with statistical fluctuation depending on the number of transits. The simulations spanned a parameters space covering disc and bulge radii between 0.2" and 2.0" and integrated V magnitudes from G=14 to G=20, regardless of their physical relevance. All simulated objects were symmetric, and 10 000 synthetic objects were generated representing of two extreme types: pure bulge brightness profiles following a de Vaucouleurs law (n=4), and pure disc profiles following an exponential law. 

The parameters adopted for the detection algorithms were the nominal ones, and thus may change after an optimization campaign. Even though optimized parameters may improve a little the detection of extended objects, significant deviations of the results presented herein are not expected. 

In figure \ref{GalaxySMdetectionsVPA} we present the resulting sky mapper detection maps for each type of object. The colormap encodes detection efficiency at each each region of the integrated magnitude vs. radius plane. This was computed as the fraction of the total simulated transits in which the objects lying at that position of the parameter space were detected. The results presented in this figure indicate that while a most de Vaucouleurs profiles were detected, there is a large detection valley among the exponential profiles. The small scale variations that can be seen overall these maps are artifacts created by the simulated Gaia scanning law and the CCD gaps in the focal plane; accordingly, they change randomly depending on the sky direction in which the simulation is performed. 

The main conclusion obtained from this test is that galaxies dominated by a disc component are expected to be very poorly detected at the sky mapper level, and thus will probably not be further confirmed by at the astrometric field level, and will not be transmitted to the ground. In contrast, most galaxies with prominent de Vaucouleurs like bulges, or elliptical galaxies, will be detected and thus probably transmitted. This corroborates the conclusions obtained in the previous sections of this work. 

\section{Distance effect}

\begin{figure*}
 \centering
 \includegraphics[width=0.82\textwidth]{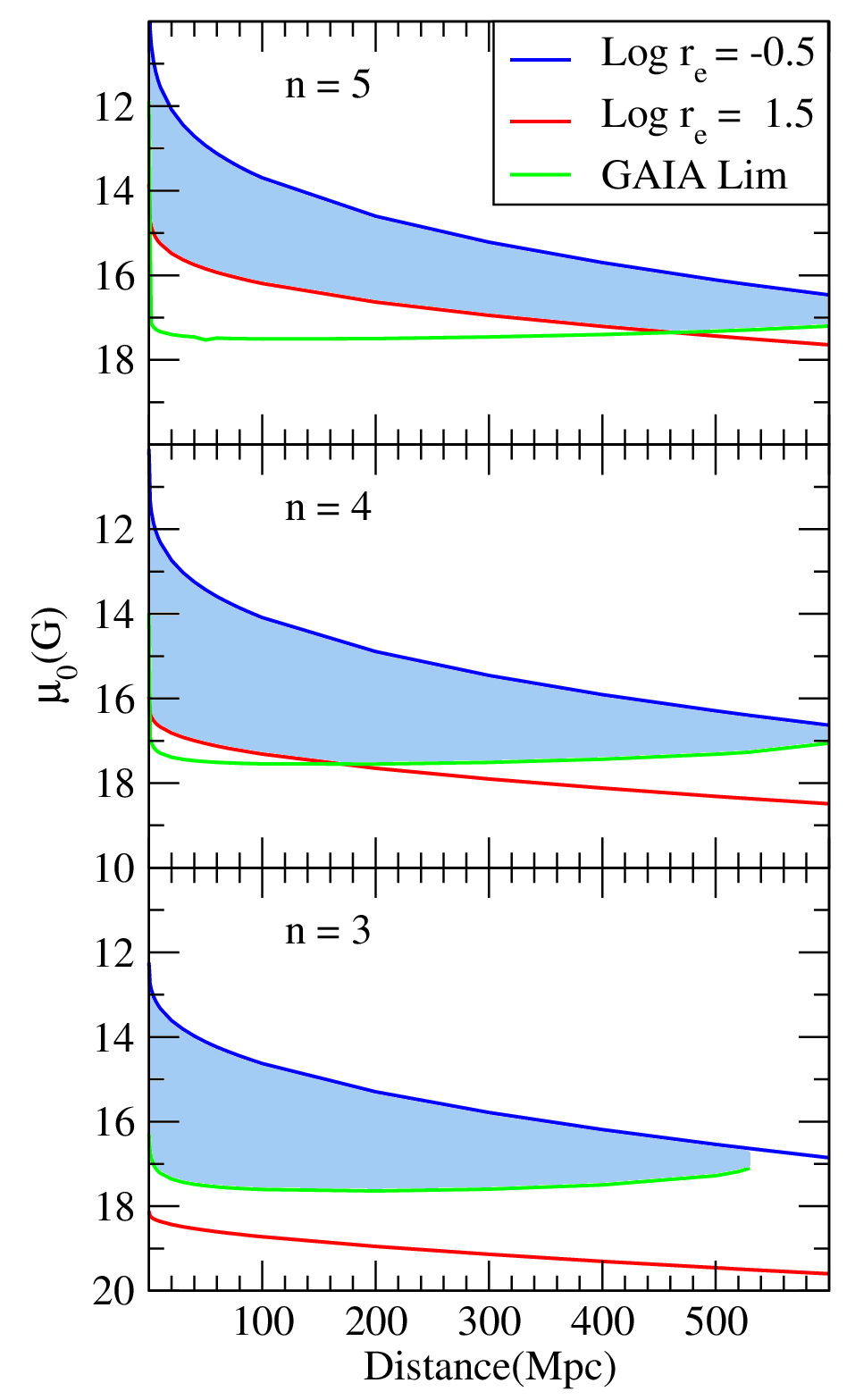}
 \caption{Effect of distance in the mean surface brightness of the central pixel of Gaia of elliptical galaxies described by S\'ersic profile. The blue and red lines corresponds to the size limits $log r_e(kpc)=-0.5$ and $log r_e(kpc)=1.5$ comprising the vast majority of known ellipticals with good profile decomposition. The expectation is that most of nearby normal, $n=4$, ellipticals in the region $d < 170$ Mpc could be detected. Their fraction will be progressively smaller up to the limit of 600 Mpc. Objects with steeper brightness profile could be detect to greater distances while those with softer profile will be limited to nearby regions.}\label{MucDistEllip}
\end{figure*}

The Gaia observations of more distant objects, even if they are close enough to be not affected by an important amount of cosmic evolution and therefore having exactly the same nuclear surface brightness as the $z=0$ objects, will be modified by two major effects. First there is a contribution of the cosmological dimming on the surface brightness due to the Tolman effect implying that $I_0(z)=I_0/(1+z)^4$. Secondly there is a geometric dilution effect since a larger spatial region will be sampled in the fixed angular size of the Gaia pixel. The result of the combination of these effects is that the corresponding average central surface brightness of the objects in the SM working windows becomes progressively fainter as we increase the distance. Consider as an example a typical relatively bright elliptical galaxy with $n=4$ and $log r_e(kpc)=0.5$ having by the Kormendy relation an effective surface brightness $\mu_e=21.33 \sbri$ at $z=0$ and $M_{abs}=-21.13$. At a distance of 10 Mpc its central surface brightness is 13.02 $\sbri$ almost identical to the corresponding value at $z=0$. At that distance we estimate by numerical integration of the S\'ersic profile that the detected magnitude of this object in the central sample of the SM detection window will be 18.03 mag while the background is 19.18 mag. After adding the counts in the 3x3 SM window, correcting for the background contribution and the Gaia PSF adopted correction we conclude that the estimated total magnitude equals to G=16.73 mag and therefore the object will be easily detected. However, using the cosmological concordance model ($H_0=70 \; {\rm km}.{\rm s}^{-1}.{\rm Mpc}^{-1}$, $\Omega_m=0.27$, $\Omega_\Lambda=0.73$) to estimate the redshift, this same object when located at 600 Mpc will have an average central surface brightness of 13.58 $\sbri$ and now the magnitude in the central sample of the SM section will be 21.20 mag while the background contribution per sample is 24.57 mag. At that distance the total corrected magnitude in the SM window will be G=19.99 mag practically equal to the stipulated limiting Gaia magnitude. A consequence of this effect is that an increasingly larger proportion of objects leave the detection region as the distance is increased and practically no objects with $n=4$ should be detected beyond 600 Mpc. This effect is shown in figure \ref{MucDistEllip} where we estimate the variation of the mean surface brightness of the detectable objects integrated in the SM central pixel of Gaia, in the euclidean approximation, as we observe objects at higher distance. The three panels illustrate the effect suffered by elliptical galaxies described by S\'ersic profiles with $n=3, 4$ and $5$. In each diagram the cyan region between the blue and red lines indicate the predicted behaviour of detected spheroids including almost the entire population of elliptical galaxies with effective radius in the interval between $log r_e(kpc)=-0.5$ and $1.5$ respectively and obeying the observed local Kormendy relation. The two extreme effective radius defining this region were choose to comprise the vast majority of ellipticals in the local Universe. The green line represent the  variation of Gaia detection limit, G=20 mag in the SM section, in the plane of the surface brightness versus increasing distance for objects satisfying the kormendy relation.

In the case $n=4$ the expectation is that even the population of ellipticals with larger effective radius, and therefore lower surface brightness and high total absolute magnitude, should be detected by Gaia up to a distance around $170$ Mpc. Above this limit we begin to loose a progressively higher fraction of ellipticals having larger effective radius. Objects with smaller effective radius tend to present higher central surface brightness and should be detected at higher distances in this nearby cosmic neighbourhood. But as the distance increase to a limit of the order of $610$ Mpc all the $n=4$ elliptical population is lost due to the adding contribution of the geometric distance dilution and the dimming due to the Tolman effect which is responsible for an increase of $0.57 \sbri$ in the central surface brightness.  The cyan shaded region indicate the fraction of the population that will have chance to be detected at each distance. An exact estimation of the extent of this effect is a hard task since we do not know exactly how real objects described by a given $n$ S\'ersic profile are spread out over the effective radius interval and moreover there is no guarantee that the S\'ersic profile could be a good description of the surface brightness in the galactic central region. The number of local objects having good photometric decomposition structure parameters is relatively small and the cosmic distribution of the effective radius among ellipticals is not precisely know. Judging from figure \ref{Mu0ReSph}, and from the homogeneous data of Caon et al (\cite{Caon1993}) and Kormendy et al (\cite{Kormendy2009}), we grossly estimate that observed objects are restricted to the indicated interval  $-0.5<log r_e(kpc)<1.5$. If we further make the arbitrary assumption that these objects are evenly spread inside this interval we can estimate that the extent of the shaded region is an indicative of the fraction of observable objects. Using this crude approximation we estimate that most of the available population of ellipticals presenting a $n=4$ profile will be observed up to a distance of 170 Mpc and approximately 1/3 of the objects between 170-600 Mpc should be lost. Above that distance corresponding to $z=0.14$ the $n=4$ ellipticals should be hardly observable by Gaia.

In the case of a $n=5$ profile presented in the superior panel of figure \ref{MucDistEllip} the indication is that the entire population of the bright nuclear elliptical will be observed in our neighbourhood. The situation for objects with softly $n=3$ profiles is more severely limited since even in our close vicinity only 50\% of them have any chance to be detected. 

One interesting quantity to obtain is the total number of probable objects to be observed. Admitting that most of the elliptical galaxies can be represented by the $n=4$ profile we can estimate in each distance interval the total number of objects. According to de Bernardi et al (\cite{Bernadi2010}) the local density number of ellipticals from the SDSS survey is estimated as $10^{-3}$ gal/Mpc$^3$. Therefore we can loosely conclude that the total number of detection by Gaia, corrected by the Milk Way zone of avoidance, should be of the order of 19 000 objects classified as elliptical up to 170 Mpc. Using these numbers we can further estimate that in the distance interval 170-600 Mpc Gaia should detect half a million of elliptical galaxies. 

The situation of bulges is even much less clear than the case of ellipticals. Quite probably the bulges of late type spirals will be difficult to detect due to their soft brightness profile. The best hope to the detection of bulges lies in the early type spirals of S0-Sa morphological classes where a class of $r^{1/4}$ spheroids are more commonly observed. Based on morphologically classified objects of the SDSS survey Nakamura et al (\cite{Nakamura2003}) conclude that the density of early type galaxies (E + S0) is of the order of $1.7 {\rm x} 10^{-3}$ gal/Mpc$^3$, for $H_0=70$ km/s/Mpc. Therefore, we might expect that the number of Gaia detections could be approximately doubled by the inclusion of the bulges of early type spiral galaxies.

\section{Conclusions} 

Based on the present status of Gaia mission our main conclusions are:

\begin{itemize}
\item It is quite unlikely that the stellar disks could be detected except if we consider the population of objects harbouring  AGN's and those presenting strong nuclear star forming episodes.

\item The major fraction of detections will be quite probably constituted by normal ellipticals and bulges of S0-Sb galaxies having steeper brightness profiles and S\'ersic indices $n=3-5$ or larger. A crude estimate indicate that the possible number of elliptical detections in the nearby 170 Mpc is situated around 19 000 objects and could be doubled by the inclusion of S0-Sa bulges. The population of very bright ellipticals present a relatively lower central surface brightness and some of them could escape detection. 

\item More distant objects, in the range 170-600 Mpc, could be observed but the evaluation of the total number of possible detections is more uncertain. If most of the population is represented by a S\'ersic profile $n=4$ then about 2/3 of them could be detected increasing the total number of source to half a million objects. 

\item Ellipticals having soft brightness profiles , $n<3$, will be more difficult to be detected and the same is valid for spheroidal galaxies.

\item This scenario is supported by numerical simulations of the detection at the sky mapper level. These simulations were performed by the Gaia Instrument and Basic Image Simulation and the video processing algorithm prototype using nominal parameters.
 
\end{itemize}

At first glance we could think that the predicted number of detections is too small in comparison with the huge expectations of Gaia in the stellar area. However, it is useful to remember that the present number of observed spheroidal objects having similar image resolution is less than one hundred objects. Therefore, if our expectations reveal to be true Gaia will provide a massive sample of new nuclear observations of ellipticals and bulges of early type galaxies in the nearby Universe. Due to the characteristics of the Gaia mission the same object will be observed several times with different scanning directions. All detected objects will have in each scanning passage a collection of the AF high resolution images as well as the lower resolution SM data. At the end this collection of data along each direction may be processed at Earth to reconstruct the estimated images. It is conceivable that some loss of resolution could result from the reconstruction process, and also that reconstruction artefacts might exist. Nevertheless it is also conceivable that the sample of Gaia observations will be of great importance for our understanding of the central structure of spheroids and their distribution in the local Universe, from its one-dimensional data alone. In the past ellipticals were considered relaxed systems that evolved passively due to little or no gas reservoir, formed from  a monolithic collapse scenario, in a single burst of intense star formation. Nowadays this picture was completely changed due to recent photometric and kinematic studies that have revealed an huge complexity in their dynamical structure, star-formation history and assembly history. These studies have been showing that ellipticals can be classified in three groups according with their luminosity. The photometric and kinematic properties of each group are very 
different and depend of their luminosity (Mo et al., \cite{Mo2010}; Blanton \& Moustakas, \cite{Blanton2009}). Therefore the high resolution  Gaia observations in the central region of ellipticals could be very useful to test these different scenarios by using an homogeneous sample of objects.

 An important fraction of the observed sample of nearby galaxies brighter than $G_{RVS}=17$ mag should also be detected in the spectrophotometer and due to its characteristics the same object will be observed several times along different angle directions. Another interesting issue is that the detected objects should also be observed in the blue (BP) and red (RP) photometers. This would be an invaluable information for the stellar population studies in the central regions of galaxies.

One interesting issue is the discussion of the scientific impact of changing the limiting detection magnitude of Gaia to $G=21$ mag. Such a change will obviously impact the quantity of data processing and is a matter under discussion by the mission control board. If this change could be consolidated we might ask on its impact in the detection of galaxies. In the case of the disks it is conceivable that this change will not modify the scenario discussed in section 3. As we can see from figure \ref{Mu0ReDisk} a variation of one magnitude in the detection limit magnitude will not be able to include the population of normal stellar disks. On the other hand that change will include a significant higher fraction of local spheroids as we can observe from figure \ref{Mu0ReSph}. Perhaps more importantly it will include a fraction of local spheroids with a soft surface brightness profile similar to those found in some of the local spheroidal ellipticals as well as a fraction of the spheroidal bulges of late spirals observed by MacArthur et al (\cite{MacArthur2003}). In the case of normal $n=4$ ellipticals the upgrade to $G=21$ mag will include in the detection list a large fraction of the objects more distant than 170 Mpc. If we consider an object with with $log r_e(kpc)=1.5$ obeying the Kormendy relation, corresponding therefore to a very luminous elliptical with $M_{abs}=-23.19$ mag, it will have at $d=600$ Mpc a total corrected magnitude in the SM window of G=21.00 mag and therefore will be include in the list of detected objects. Therefore the limiting distance for completion will triplicate expanding our number estimate approximately by a factor of 30.

\begin{acknowledgements}
Part of this work was partially supported by CAPES/COFECUB and by the Portuguese agency FCT (SFRH/BPD/74697/2010, PTDC/CTE-SPA/118692/2010). We also thank the French CNES for providing computational resources, and the Gaia Data Processing and Analysis Consortium Coordination Unit 2 - Simulations for the {\it Gaia Instrument and Basic Image Simulator}. We also acknowledge Jos de Bruijne for kindly providing information regarding the functioning of the onboard VPA detection software. We also thanks an anonymous referee for the helpful comments.
\end{acknowledgements}

\end{document}